\documentclass[prl,twocolumn,floatfix,showpacs,superscriptaddress]{revtex4}
\usepackage{graphics}
\usepackage{dcolumn}
\usepackage{bm}
\usepackage{epsfig}
\usepackage{amsmath}
\usepackage{amssymb}

\begin{document}
\draft
\preprint{HEP/123-qed}
\title{Non-local network-excitation 
in the Ising antiferromagnet on the anisotropic triangular lattice}
\author{Chisa Hotta}
\affiliation{Kyoto Sangyo University, Department of Physics, Faculty of Science, Kyoto 603-8555, Japan}
\author{Tetsuhiro Kiyota}
\address{Aoyama-Gakuin University, 5-10-1, Fuchinobe, Sagamihara, Kanagawa 229-8558, Japan}
\author{Nobuo Furukawa}
\address{Aoyama-Gakuin University, 5-10-1, Fuchinobe, Sagamihara, Kanagawa 229-8558, Japan}
\address{Multiferroics Project (MF), ERATO,
  Japan Science and Technology Agency (JST), The University of Tokyo, Tokyo 113-8656, Japan}
\date{submitted in October 12, 2009}
\begin{abstract}
We study the Ising antiferromagnet on the triangular lattice which are interacting 
along three different directions via two $J$-bonds and one anisotropic $J'$-bond with $J' \ge J$. 
Although its finite temperature state has long been considered as simply 
disordered, we find a systematic generation of 
a ferromagnetic $J'$-bonds named "good defects" each carrying the energy $2\delta_J\equiv 2(J'-J)$. 
They exhibit an eminent correlation which is regarded as a network or a soft lattice. 
The specific heat shows a universal peak at $T^* \sim \delta_J$ and 
the corresponding energy follows the activation type of temperature-dependence 
with an excitation gap $2\delta_J$. 
Energetics and the crossover between this particular two-dimensional state 
and the other disordered or one-dimensional states are discussed. 
\end{abstract}
\pacs{75.40.-s, 75.50.Mm, 72.80.Ng, 72.80.Le}
\maketitle
\narrowtext
%
The Ising model continues to be the starting point of understanding the basic concept of magnetism 
from the date of its proposal\cite{ising25}. 
In the midth of the last century, the exact solutions by Onsager\cite{onsager44}, Wannier\cite{wannier45}, 
and others\cite{kano53,houtappel50} successively clarified its nature on the two dimensional (2D) lattices. 
Particularly, the phase transition at finite temperature found on the square lattice\cite{onsager44} 
remains a prototype description of the conventional phase transition and critical phenomena. 
Meanwhile, still at present, there are representative models whose low energy properties are studied extensively, 
some of them are related to a series called the frustrated systems\cite{diep}. 
\begin{figure}[tbp]
\begin{center}
\includegraphics[width=8cm]{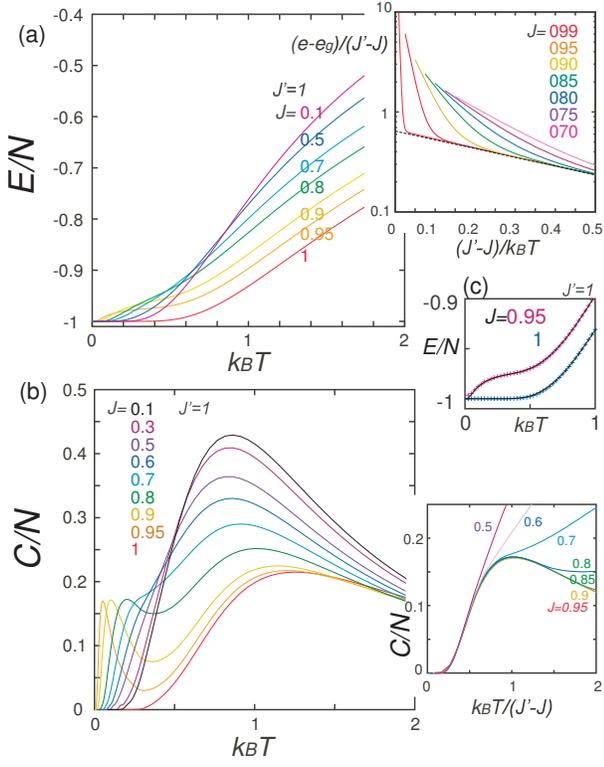}
\end{center}
\caption{
Panels (a) and (b) are the energy per site, $e=E/N$, and the specific heat, $c={\rm d}e/{\rm d}T$, 
as a function of temperature $T$ for various choices of $J \leqq J'=1$. 
The inset of panel (a) gives an Arrhenius plot for 
$(e-e_{\rm g})/\delta_J$ as a function of scaled temperature $T/\delta_J$. 
Solid line gives the activation-function with $\Delta=2\delta_J$. 
Inset of panel (b) gives the replot of $c$ as a function of $k_BT/\delta_J$. 
Panel (c) shows the direct comparison of energy 
the Monte Carlo simulation (marks) and the exact solution (solid lines) at $J=1$ and  $0.95$. 
}
\label{f1}
\end{figure}
%
\par
Triangular lattice systems are the representative geometrically frustrated systems 
which are studied in the context of the SU(2)\cite{lhuillier92,sorella99,misguish99}, 
anisotropic \cite{kawamura85,miyashita93,moessner01} spin models, 
and in electronic systems as well which yields a spin liquid Mott insulator\cite{imada02,yoshioka09}. 
While the ground state of these models are quite often disputing, 
the low energy excitations of such frustrated systems are far more difficult, 
where we expect the dense spectrum due to the many-body effect. 
One of the clue to understand them is to start from the Ising limit and 
include quantum or thermal fluctuations\cite{ch-furukawa,chisa-f08}. 
The present Letter actually starts from the exact solutions of 
the triangular Ising antiferromagnet\cite{wannier45} which is classically disordered 
throughout the whole temperature range. 
Revisiting this "already well understood" system, 
we clarify a new aspect of thermal fluctuation in the anisotropic but still heavily frustrated region. 
The Hamiltonian is given as, 
\begin{equation}
{\cal H} = \sum_{\langle i,j \rangle} J_{ij} \sigma_i \sigma_j 
\end{equation}
where $\sigma_i=\pm 1$ and interactions are confined to nearest neighbor sites only. 
Throughout this Letter we focus on the case where the interactions along bonds in one direction 
$J_{ij}=J'$, which is taken as a unit of energy, 
are larger than in the other two directions, $J_{ij}=J$, namely $0 < J < J'$. 
Hereafter we regard the direction along the strong $J'$-bonds as the chain direction. 
The ground state at $J<J'$ is a combination of one-dimensional (1D) N\'eel order 
along the strong $J'$ chains and a disorder in the remaining two directions of $J$-bonds 
where the spin-spin correlation along decay in power law. 
We find that at finite temperature, the partial disorder embedded in this system couples with the 
thermal fluctuation and yields an exotic state where the ferromagnetic $J'$-bonds 
form soft non-local quasi-2D network. 
\par
We first analyze thermodynamical properties obtained by the exact solution\cite{houtappel50}. 
Figures~\ref{f1}(a) and ~\ref{f1}(b) present the exact energy $e=E/N$ and 
the specific heat $c={\rm d}e/{\rm d}T$ for various values of $J\leqq 1$. 
There is no discontinuity in these bulk properties. 
We find, however, a peak or its precursor of the specific heat at low temperatures when $J>0.5J'$. 
The inset of Fig.~\ref{f1}(a) shows an Arrhenius plot of the dimensionless energy (with respect 
to the ground state energy $e_{\rm g}=-J'$) in unit of $\delta_J \equiv J'-J$ as a function of dimensionless temperature. 
We find that these curves asymptotically approach an activation-type behavior with the energy gap 
$\Delta= 2\delta_J$. This characteristic energy scale is also found in the universal behavior of 
the specific heat in the inset of Fig.~\ref{f1}(b) as a function of normalized temperature, 
$T/\delta_J$, where the peak at $T/\delta_J \sim 1$ is scaled for all cases of $0.5<J<1$. 
\par
\begin{figure}[tbp]
\begin{center}
\includegraphics[width=8cm]{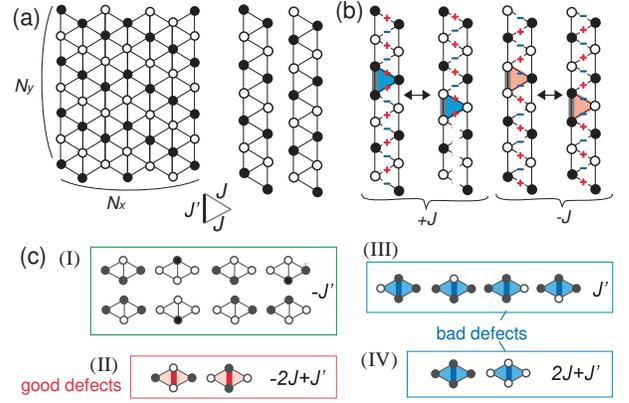}
\end{center}
\vspace{-5mm}
\caption{
Configurations of (a) ground state and (b) a single defect state. 
Black and white circles represent the up and down spins, respectively. 
The $J'$-bond which connects identical spins
are called "defects". 
(c) Classification of the unit plaquettes according to their energy. See main text for details.
}
\label{f2}
\end{figure}

Let us examine the possible configuration of spins at low energy 
in order to understand the origin of the characteristic energy scale $\Delta$. 
We start from the disordered ground state as shown in Fig.~\ref{f2}(a). 
Here, each chain has a N\'eel order.
Inter-chain interactions $J$ are cancel out on the whole
irrespective of a relative phase of antiferromagnetic spin configurations between the chains, 
as shown in the right panels of Fig.~\ref{f2}(a), and thus the ground state
has degeneracy.
An excited state is generated by introducing a ferromagnetic configuration on a $J'$-bond as 
illustrated in Fig. \ref{f2}(b), which we hereafter refer to as a "defect". 
Each defect takes an excitation energy of $2J'$ along the chain direction, 
while between the chains it creates an uncanceled inter-chain interaction energy.
In order to examine the energetics of the defects in detail, we decompose the
lattice into plaquettes depicted in Fig.~\ref{f2}(c),
and count the energy of them by summing up the energy of $J'$- and $J$-bonds
within the plaquette. Here, we take half of the energy of the four $J$-bonds on the edges 
of the plaquettes which are shared by the neighbors. 
Plaquettes are energetically classified into four types. Those without defects 
are assigned (I) with an energy $-J'$, where the interchain interactions are indeed canceled out. 
Each plaquette categorized in (II) has a defect 
which gains the interchain bond energy, and the energy amounts to $J'-2J$. 
Those classified as (III) and (IV) have higher total energies, $J'$ and $J'+2J$, respectively. 
Thus, in the region $J<J'$, plaquettes (I) have the lowest energy, while
plaquettes (II) have the first excited state energy which we hereafter call as "good" defects, 
and the plaquettes (III) and (IV) will be contrastingly referred to as  "bad" defects. 
The excitation energy of defect (II) actually correspond to the characteristic 
energy scale $\Delta=2\delta_J$ in the previous exact solution. 
If $\delta_J$ is sufficiently small, 
the low energy excitations by the bad defects can be negligible 
and the state with only "good" defects may dominate. 

\par
In order to examine to what extent the good defects dominate at low temperatures,
we perform the classical Monte Carlo simulation (MC).
Calculations are carried out mainly on a $N=N_x\times N_y = 50\times 100$ site cluster 
up to the observation time of 5000 Monte Carlo steps (MCS) after the relaxation 
(which takes less than 500 MCS). We take about 200 independent runs for averaging. 
For each snap shot we separately count the number of good defects ($N_g$) from 
the total number of defects ($N_d$). 
We confirm that the energy of MC almost exactly coincides with that of the exact solutions as 
in the inset of Fig.~\ref{f1}(c) by the dotted (MC) and solid(exact) lines. 
\par
\begin{figure}[tbp]
\begin{center}
\includegraphics[width=8.5cm]{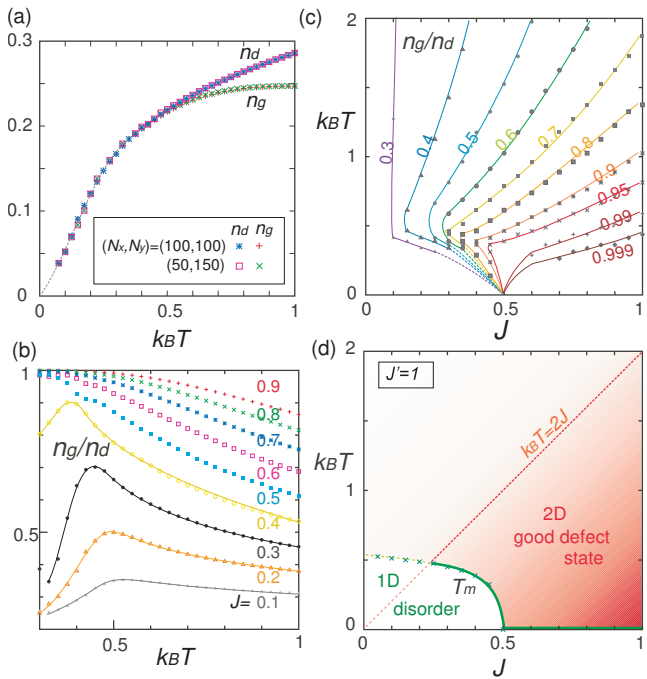}
\end{center}
\caption{
The results of the Monte Carlo simulations at $J'=1$. 
(a) Density of defect(II-IV), $n_d$, and good defects, $n_g$, 
for different sizes of clusters as a function of temperature at $J'=1$. 
(b) $n_g/n_d$ as a function of temperature.
(c) Contour plot of $n_g/n_d$ by $J$ and $k_BT$. 
(d) $J$-$T$ diagram which depict the dimensional crossover. 
}
\label{f3}
\end{figure}

We first present the $T$-dependence of $n_d\equiv N_d/N$ and $n_g\equiv N_g/N$ 
at $J=0.95$ in Fig.~\ref{f3}(a) for various cluster sizes. 
We confirm that size effects are negligible here.
Both $n_d$ and $n_g$ decreases exponentially as $T$ is lowered.
At higher temperature, the number of bad defects, $n_b=n_d-n_g$, increases rapidly. 
A distinct feature is found in the $T$-dependence of $n_g/n_d$ in Fig.~\ref{f3}(b) 
at around the particular degree of anisotropy, $J=0.5$; 
when $J>0.5$, the ratio $n_g/n_d$ increases monotonically toward unity with decreasing temperature, 
whereas at $J<0.5$, a peak exists at $T=T_m$. 
The behavior of $n_g/n_d$ is better understood in its contour plot 
on the $J$-$T$ diagram given in Fig.~\ref{f3}(c).
Now, we plot $T_m$ in a $J$-$T$ diagram in Fig.~\ref{f3}(d). 
In the shaded region the good defects are energetically 
chosen to be thermally activated. 
Here, $k_BT=2J$ which is the energy difference between bad (III) and good (II) defects, 
approximately gives the crossover line which characterizes the temperature 
that the bad defect is activated. 
Above this crossover line, the difference between the good and bad defects no more makes sense. 
Rather, any defects are activated randomly as the temperature is increased. 
\par
%
\begin{figure}[tbp]
\begin{center}
\includegraphics[width=8cm]{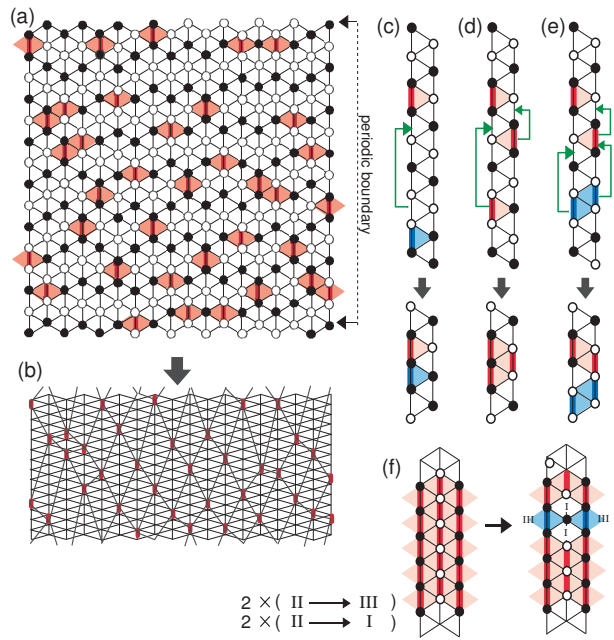}
\end{center}
\caption{
(a) Representative configuration of the "good defect state" with $n_g/n_d=1$. 
Arrows indicate the two sites identical under the periodic boundary condition along the chains. 
(b) Schematic "soft lattice structure" of good defects corresponding to (a). 
Panels (c), (d) and (e) describes the multiple defect configuration embedded 
in the antiferromagnetic background. See main text for details. 
The lower panels of each (c)-(e) show the equivalent case when 
the defects are shifted and squeezed along the chain.
(f) The cases where the two defects on the center chain are depleted. 
}
\label{f4}
\end{figure}
\par
Let us next discuss the nature of the good defect states in detail.
We find in the MC results that defect-defect correlation decays quickly toward zero after a few
chain distance in the interchain-direction (not shown). 
Then, the defects seemingly distribute randomly in real space, 
which is however, not the case. 
Instead, an distinguished correlation arises in the positions of good defects
as a consequence of eliminating bad defects.
\par
One finds the following configuration rule of the good defect state; 
{\it to place the good defects exclusively,
 those on two neighboring chains must always be located alternatively 
along the chain.} 
We indeed confirm in the MC that all the snap shots with $n_g/n_d=1$ 
strictly follows this rule. 
The representative good defect configuration is given in Fig.~\ref{f4}(a). 
Here, all the defects are good ones, and if we neglect the background spins 
as in Fig.~\ref{f4}(b), the good defects form soft lattice structure. 
Here, we mean by "soft" that the absolute location of defects do not matter, 
i.e., defects are shifted in the chain direction without breaking the above rule only 
if it does not go over the defects on the neighboring chains. 
\par
We now explain the energetics behind the above rule step by step. 
It is noteworthy that the relative phase of the N\'{e}el state 
between the neighboring chains essentially determines 
the energy of a defect when it is embedded as shown in Fig.~\ref{f2}(b). 
We start from this figure noting that the position of the defect can be shifted translationally 
along the chain without changing its energy. 
This is because such a shift by unit length flips 
all the spins in a defect-plaquette, in which case a good(bad) defect remains good(bad) 
with the energy unchanged. 
Next, we consider multiple defects on a chain.
Note that if we restrict ourselves to the periodic boundary condition with even number of sites 
along the chain, the defects should always appear as pairs. 
Figure \ref{f4}(c) shows two defects on a single chain. 
A pair of defects on isolated chain is always a combination of 
good and bad ones. 
This is easily understood if the defects are shifted and 
placed next to each other as in the lower panel of Fig.~\ref{f4}(c). 
Thus, as long as we consider pairs of defects on an isolated chain 
the bad defects cannot be excluded. 
Now, in order to make both defects good, one need to twist the phase of 
antiferromagnetic configuration on the neighboring chain. 
This corresponds to inserting one defect into the neighboring chain between the previous 
two as in Fig.~\ref{f4}(d), and the introduced defect also turns out to be a good one. 
If we insert one more defect as in Fig.~\ref{f4}(e), 
we again cannot exclude bad defects, since two consecutive defects 
bring the phase of the antiferromagnetic configuration back. 
The picture is easily extended to multiple chains, and the configuration pattern 
in Fig.~\ref{f4}(d) is extended to that of Fig.~\ref{f4}(a). 
This straightforwardly leads to the "good defect rule" we mentioned earlier. 
\par
The remaining issue is the crossover at $J=0.5$, which is 
related to the stability of the soft lattice state. 
Consider depleting a pair of defects on one of the chains in this state. 
The equivalent case where a good defects are packed closest together is given in Fig.~\ref{f4}(f). 
The depletion corresponds to replacing two (II)-plaquettes on the chain by (I) and 
two (II)-plaquettes in the neighboring chains in between the depleted two by (III), 
and the net energy gain becomes $dE=2\times (2J-J')$. 
At $J<0.5J'$, we have $dE <0$ and the soft lattice is energetically unstable. 
Then, the isolated defects shown in Fig.~\ref{f4}(c) can be realized at low temperatures, 
namely at $T<T_m$ and $J<0.5$ in Fig.~\ref{f3}(d). 
Note that this configuration of defects corresponds to the classical version of fractional charges 
in the systems with quantum fluctuation\cite{chisa-f08}. 
\par
In the presence of the soft lattice structure,
density fluctuations of defects show characteristic behaviors; 
it is disfavored to change numbers of defects independently 
on each chain since such fluctuations destroy the soft lattice structure. 
Instead, defects must be inserted/depleted simultaneously on all chains in such a way
that the good defect rule is maintained. 
Actually, in MC when an additional defect is introduced on a certain chain, 
it continuously spreads to the neighboring chain 
to form a soft domain wall or a soft dislocation line
in the interchain direction to make all chains finally equivalent. 
This nature of the defect density fluctuations changes drastically 
once we go over the crossover line at $J=0.5$. 
At $J<0.5$, it is energetically allowed to change the defect densities
on each individual chain independently. 
Namely, we may regard that the defect density fluctuation have 1D characters, 
similar to that found in the limit $J=0$.
\par
The $J$-$T$ diagram in Fig.~\ref{f3}(d) is thus well classified by the dimensional crossover lines.
In the whole region of $0 \leqq J < 1$, 
spins in the ground state exhibit a 1D N\'{e}el configurations along the chain, 
and have degeneracy in such a way that all the spins on a single chain can be flipped 
simultaneously without any cost of energy. 
Namely, the interchain coupling $J$ is irrelevant.
However, once we enter finite temperature region at $J>0.5$ 
where the 2D soft lattice network of good defects is present, 
such 1D fluctuation disappears.
This is because if we flip all the spins simultaneously on one chain 
in the soft lattice state, the good defects on this chain and on the neighboring 
two chains are all transformed to bad ones, and the energy increases. 
In other words, the relative phases between the chains are locked 
in forming the soft lattice state. 
Then, instead of the above mentioned 1D fluctuations, 
the 2D defect density fluctuations appear, 
which allows for the change of defect number on all chains simultaneously. 
The ground state at $J=0.5J'$ is thus a singular point 
where a dimensional crossover between $T=0$ and $T>0$ emerges.
\par
To summarize we studied the triangular Ising antiferromagnet interacting 
with one strong anisotropic $J'$- and two weaker $J$-bonds. 
The system systematically generates ferromagnetic $J'$-bonds (defects) under thermal fluctuation, 
which exhibit an eminant non-local correlation throughout the system by conserving their number 
and by keeping their relative locations. 
This state which is regarded as two-dimensional soft network structure, 
cannot be described by the conventional "long range order", 
since it is neither characterized by the spin-spin  nor the defect-defect 
correlation functions. 
The present findings will provide clue to understand the low-energy excitation of the 
anisotropic triangular antiferromagnet under fluctuations 
from the Ising limit\cite{ch-furukawa,chisa-f08,kohno}. 
As far as we know, such novel phenomena has never been found, 
and shall provide a new example in exploring the scheme beyond the conventional statistical or 
condensed matter physics. 
\par
We thank Y. Motome and Kenn Kubo for discussions. 
This work was partially supported by Grant-in-Aid for Scientific Research from the MEXT, Japan. 

\end{document}